\documentclass[useAMS,usenatbib,article]{mn2e}
\usepackage[dvips]{graphicx}

\def\lesssim{\mathrel{\hbox{\rlap{\hbox{\lower4pt\hbox{$\sim$}}}\hbox{$<$}}}}
\def\gtrsim{\mathrel{\hbox{\rlap{\hbox{\lower4pt\hbox{$\sim$}}}\hbox{$>$}}}}
\newcommand{\mincir}{\raise
-2.truept\hbox{\rlap{\hbox{$\sim$}}\raise5.truept
\hbox{$<$}\ }}
\newcommand{\magcir}{\raise
-2.truept\hbox{\rlap{\hbox{$\sim$}}\raise5.truept
\hbox{$>$}\ }}
\newcommand{\be}{\begin{equation}}
\newcommand{\ee}{\end{equation}}
\newcommand{\ba}{\begin{eqnarray}}
\newcommand{\ea}{\end{eqnarray}}

\title[NIR Luminosity and Stellar Mass Functions]{ACCESS: NIR
  Luminosity Function and Stellar Mass Function of Galaxies in the
  Shapley Supercluster Environment\footnote{Based on observations
  collected at UKIRT with the WFCAM}}

\author[P. Merluzzi et al.]{P. Merluzzi$^{1}$, A. Mercurio$^{1}$, C.P.
  Haines$^{2}$, R. J. Smith$^{3}$, G. Busarello$^{1}$,
  J. R. Lucey$^{3}$\\ $^1$INAF-Osservatorio Astronomico di
  Capodimonte, I-80131 Napoli\\ $^2$ School of Physics and Astronomy,
  University of Birmingham, Birmingham B15 2TT\\ $^3$ Department of
  Physics, University of Durham, Durham DH1 3LE}

\begin{document}

\date{Accepted . Received }

\pagerange{\pageref{firstpage}--\pageref{lastpage}} \pubyear{}

\maketitle

\label{firstpage}

\begin{abstract}
  We present the near-infrared luminosity and stellar mass functions
  of galaxies in the core of the Shapley supercluster at $z$=0.048,
  based on new $K$-band observations carried out at the United Kingdom
  Infra-Red Telescope with the Wide Field Infrared Camera in
  conjunction with $B$- and $R$-band photometry from the Shapley
  Optical Survey, and including a subsample ($\sim$650 galaxies) of
  spectroscopically confirmed supercluster members. These data sets
  allow us to investigate the supercluster galaxy population down to
  M$_K^\star$+6 and $\mathcal{M}$=10$^{8.75}$M$_\odot$. For the
  overall 3\,deg$^2$ field the $K$-band luminosity function (LF) is
  described by a Schechter function with M$_K^\star$=--24.96$\pm$0.10
  and $\alpha$=--1.42$\pm$0.03, a significantly steeper faint-end
  slope than that observed in field regions. We investigate the effect
  of environment by deriving the LF in three regions selected
  according to the local galaxy density, and observe a significant
  ($2\sigma$) increase in the faint-end slope going from the high-
  ($\alpha$=--1.33) to the low-density ($\alpha$=--1.49) environments,
  while a faint-end upturn at M$_{K}>$--21 becomes increasingly
  apparent in the lower density regions.  The galaxy stellar mass
  function (SMF) is fitted well by a Schechter function with
  log$_{10}$($\mathcal{M}^\star$)=11.16$\pm$0.04 and
  $\alpha$=--1.20$\pm0.02$. The SMF of supercluster galaxies is also
  characterised by an excess of massive galaxies that are associated
  to the brightest cluster galaxies. While the value of
  $\mathcal{M}^*$ depends on environment increasing by 0.2 dex from
  low- to high-density regions, the slope of the galaxy SMF does not
  vary with the environment. By comparing our findings with
  cosmological simulations, we conclude that the environmental
  dependences of the LF are not primary due to variations in the
  merging histories, but to processes which are not treated in the
  semi-analytical models, such as tidal stripping or harassment. In
  field regions the SMF shows a sharp upturn below
  $\mathcal{M}$=10$^{9}$M$_\odot$, close to our mass limit, suggesting
  that the upturns seen in our $K$-band LFs, but not in the SMF, are
  due to this dwarf population. The environmental variations seen in
  the faint-end of the $K$-band LF suggests that these dwarf galaxies,
  which are easier to strip than their more massive counterparts, are
  affected by tidal/gas stripping upon entering the supercluster
  environment.

\end{abstract}


\section{Introduction}
\label{intro}
The properties and evolution of galaxies are strongly related to their
environment (e.g. Blanton et al. \citeyear{bla05a}; Rines et
al. \citeyear{rin05}; Baldry et al. \citeyear{BBB06}), through the
mass and merging histories of their host dark matter halos, and the
impact of different physical mechanisms (e.g. Treu et
al. \citeyear{tre03}) that are linked in various ways to the local
galaxy density and the properties of the intergalactic medium.  In the
local Universe this environmental dependence has been investigated and
observed in the distribution of galaxy luminosities and stellar
masses, providing constraints on the assembly of galaxies over cosmic
time (see below).

Since the NIR light is dominated by established old stellar
populations rather than by recent star-formation activity, the NIR LF
can be considered as reliable estimator of the stellar mass function
(SMF, Gavazzi et al. \citeyear{GPB96}; Bell \& de Jong
\citeyear{BdJ01}) and the shape of the NIR LF constrains the scenarios
of galaxy formation (e.g. Benson et al. \citeyear{BBF03}), with
different energetic feedback processes from supernovae and AGN
required to simultaneously fit the LFs at the faint and bright ends
respectively \citep{BBM06}. Early versions of the hierarchical galaxy
formation model predicted a decrease of the abundance of massive
galaxies with redshift (Kauffmann \& Charlot \citeyear{KC98}) which
have continued forming until recent times through processes of merging
and accretion, and a steep mass function due to the presence of a
large number of faint dwarf galaxies witnessing the small dark halo
formation in the early universe (e.g. Kauffmann et
al. \citeyear{KWG93}).

The NIR LFs observed at different redshifts turn out to be well
described by Schechter functions, although there is some evidence for
an excess of bright galaxies with respect to the best fitting
Schechter function (e.g. Jones et al. \citeyear{JPC06}), while a
faint-end upturn has also been observed in some cases (e.g. De Propris
\& Pritchet \citeyear{DPP98}; Balogh et al. \citeyear{BCZ01}; Jenkins
et al. \citeyear{JHM07}).  The results supporting the hierarchical
scenario (e.g. Kauffmann \& Charlot \citeyear{KC98}) are contrasted by
other works that bring it into question (Kodama \& Bower
\citeyear{KB03}; De Propris et al. \citeyear{DSE99}; \citeyear{DSE07})
or cannot interpret univocally their findings (e.g. Drory et
al. \citeyear{DBF03}). For instance, there is a consensus that the
evolution of the characteristic absolute luminosity M$^\star$ for both
field and cluster galaxies can be described by a passively evolving
population formed in a single burst at redshift $z$=1.5-2 (e.g. Lin et
al. \citeyear{LMG06}; Wake et al. \citeyear{WNE06}; De Propris et
al. \citeyear{DSE99}; \citeyear{DSE07}). On the other hand, Drory et
al. (\citeyear{DBF03}) attribute the observed evolution in the
$K$-band LF of cluster galaxies, either to a change in the
mass-to-light ratio alone (i.e. passive evolution), or to a
combination of changes in M/L and stellar mass which could be due to
star formation and/or to merging or accretion. Recent semi-analytic
models incorporating AGN feedback have been able to reproduce well the
evolution of the SMF over $0{<}z{<}5$ (Bower et al. \citeyear{BBM06}),
predicting the observed population of massive galaxies at
$z{>}2$. However, in order to properly use the NIR, LF to disentangle
between possible scenarios of galaxy evolution, one has to consider
the dependence of the LF on environment, galaxy colour, spectral type
and morphology.

The LF of field galaxies at NIR wavebands has been primarily
investigated through the Two Micron All Sky
Survey\footnote{http://www.ipac.caltech.edu/2mass/} (2MASS) often
complemented by either optical photometry from the Sloan Digital Sky
Survey\footnote{http://www.sdss.org/} (SDSS) or wide-field
spectroscopic surveys such as the 2dF Galaxy Redshift
Survey\footnote{http://www.mso.anu.edu.au/2dFGRS/} (2dFGRS). These
works (e.g. Kochanek et al. \citeyear{KPF01}; Cole et
al. \citeyear{CNB01}) agree in describing the NIR LFs with Schechter
functions characterised by a rather flat faint-end slope $\alpha$ from
--0.77 to --0.96 and which are independent of the morphological type
(Kochanek et al. \citeyear{KPF01}), in contrast with that found in
optical surveys where the faint-end slope is steeper for late-type
galaxies. On the contrary, Bell et al. (\citeyear{BMK03}) found that
NIR LF has a brighter characteristic luminosity and shallower slope
for early-type galaxies with respect to the later types. More
recently, Jones et al. (\citeyear{JPC06}), using the 6dF Galaxy
Redshift Survey\footnote{http://www.aao.gov.au/local/www/6df/}
(6dFGRS, Jones et al. \citeyear{JSC04}) derived NIR LFs for field
galaxies 1-2\,mag deeper in absolute magnitude with respect to the
previous works. They found that the Schechter function is not ideal to
reproduce the data since it cannot match the bright- and faint-end
simultaneously, due to an excess of galaxies at magnitudes brighter
than M$^\star$.

\begin{figure*}
\centerline{\resizebox{17.5cm}{!}{\includegraphics{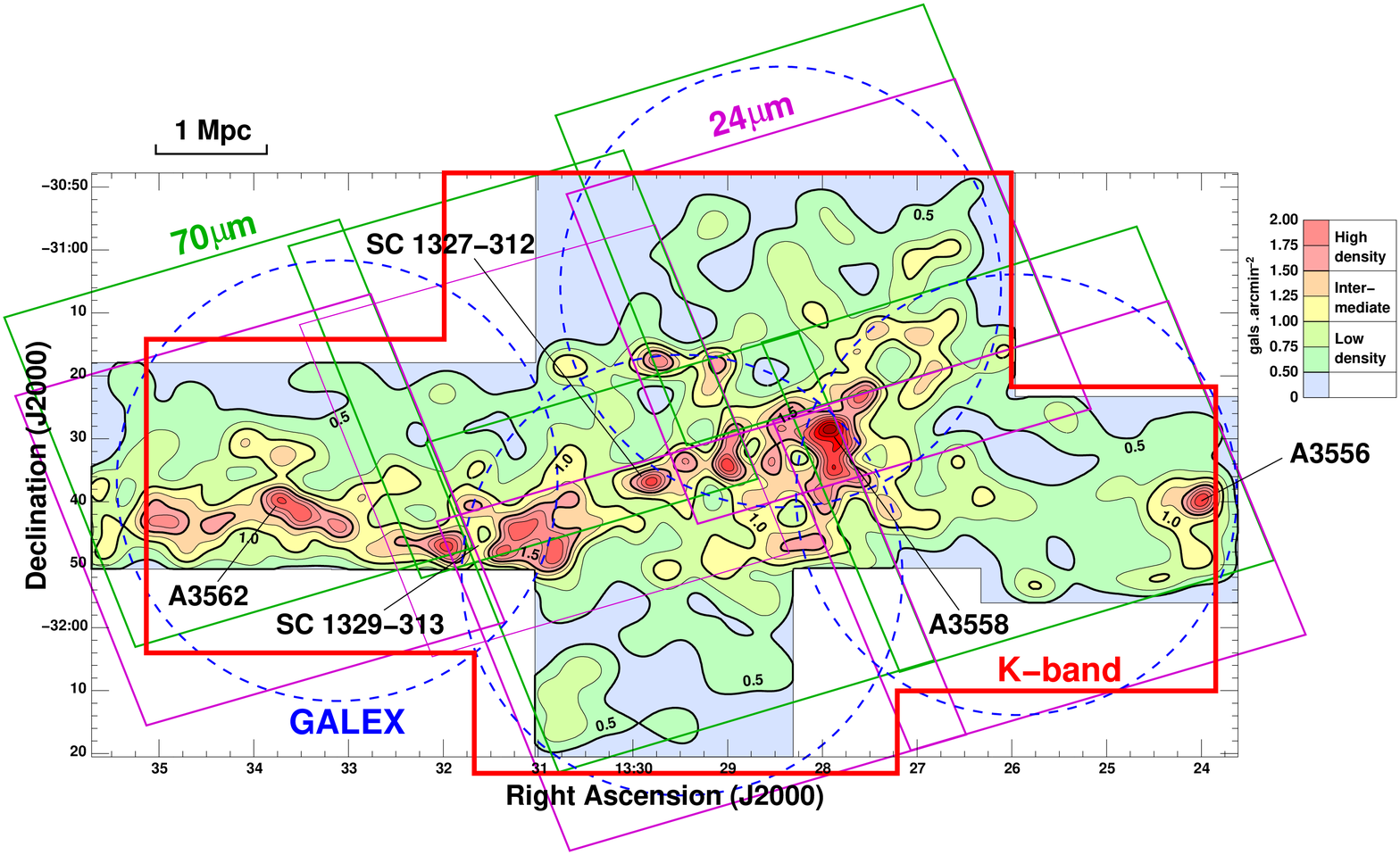}}}
\caption{The Shapley NIR survey (thick red) is shown superimposed to
the surface density of $\mathrm{R}<21.0$ galaxies as derived from the
SOS (see text). Isodensity contours are shown at intervals of 0.25
galaxies arcmin$^{-2}$, with the thick contours corresponding to 0.5,
1.0 and 1.5 galaxies arcmin$^{-2}$, the densities used to separate the
three cluster environments. The multi-wavelength photometric coverage
of ACCESS on the SSC are also shown. Magenta: 24\,$\mu$m, green:
70\,$\mu$m, blue dashed: 150\,nm and 250\,nm.}
\label{fig1}
\end{figure*}

Balogh et al. (\citeyear{BCZ01}) investigated the dependence of the
infrared galaxy luminosity function and the associated galaxy SMF on
environment and spectral type by means of 2MASS and Las Campanas
Redshift Survey (LCRS, Shectman et al. \citeyear{SLO96}) for galaxies
brighter that M$_J$=--19\,mag. In field environments the LF of
galaxies with emission lines turns out to have a much steeper
faint-end slope ($\alpha$=--1.39) compared to that of galaxies without
emission lines ($\alpha$=--0.59). On the other hand, in the cluster
environment, even the non-emission line galaxies have a steep
faint-end LF ($\alpha$=--1.22). This difference is almost entirely due
to the non-emission line galaxies which dominate the cluster
population, and present a slope close to that of the overall
field. Thus, they suggested that the cluster population is built up by
accreting field galaxies with little effect other than the cessation
of star formation. Differences in the shape of the LF for late- and
early-type cluster galaxies has been found by Huang et
al. (\citeyear{HGC03}): the late-type galaxies having a systematically
fainter M* and steeper faint-end slope. A possible faint-end upturn in
the $H$-band LF was already suggested by De Propris et
al. (\citeyear{DES98}, see also Andreon \& Pell\'o \citeyear{AP00})
for the Coma cluster outlining the steep trend of dwarf galaxies down
to M*+5, even if they did not provide a precise estimate of the
faint-end slope because of possible field contaminations. An increase
of the faint-end slope of Coma was recently observed at 3.6\,$\mu$m by
Jenkins et al. (\citeyear{JHM07}) indicating a large number of faint
red galaxies. However, Rines \& Geller (\citeyear{rin08}) found no such
upturn for Virgo, based on a fully spectroscopically confirmed sample,
and suggested that many of the photometrically-selected red sequence
galaxies which contribute to the upturns seen in other clusters are
background galaxies.

Finally, the tight correlation between the total galaxy NIR luminosity
and the cluster binding mass (Lin et al. \citeyear{LMS03},
\citeyear{LMS04}; Ramella et al. \citeyear{RBG04}) allows to probe
that the global cluster $K$-band mass-to-light ratio decreases with
cluster radius (Rines et al. \citeyear{RGD04}) showing that the
environment affects the shape of the LF also within the clusters.

We note that most of the previous works are based on the 2MASS data
which has a detection sensitivity (10$\sigma$) of $K$=13.1\,mag for
extended sources (Cole et al. \citeyear{CNB01}, $K$=13.57\,mag
according to Bell et al. \citeyear{BMK03}), limiting studies of the
environmental impact on the NIR LF to only a sample of local
clusters (e.g. Virgo), or limiting to magnitudes $<$M$^\star$+2 (e.g. Rines et
al. \citeyear{RGD04}). However, the dominant processes that quench
star-formation, and therefore transform galaxies, depend crucially on
the galaxy mass (e.g. Haines et al.  \citeyear{HLM06};
\citeyear{HGL07}), and the strong bimodality in the properties of
galaxies about a characteristic stellar mass of $\sim 3\times 10^{10}
M_{\odot}$ ($\sim$ M*+1, Kauffmann et al.  \citeyear{KHW03}) implies
fundamental differences in the formation and evolution of giant and
dwarf galaxies (e.g. Dekel \& Birnboim \citeyear{DB06}; Kere\^s et
al. \citeyear{KKW05}). This issue needs data-sets reaching much fainter
luminosities than those of 2MASS to be investigated in order to
obtain, in general, an overall picture of galaxy evolution and, in
particular, to establish the contribution of the dwarf galaxy
population to the total stellar mass in the local universe and the
physical origin of the claimed faint-end upturn. The recent
development of wide-field NIR imagers on 4-m class telescopes such as
UKIRT/WFCAM and KPNO/NEWFIRM has opened the possibility of NIR surveys
to be $>100\times$ more sensitive covering many square degrees
(e.g. UKIDSS), allowing the $K$-band LF of nearby clusters to be
obtained covering not only the cluster cores, but the entire
virialized regions.

In this context we study the $K$-band LF of the Shapley supercluster
core (SSC) down to the dwarf regime (reaching $\sim$M$^\star$+6) with
the aim of i) quantifying the environmental impact on the shape of the
NIR LF; ii) deriving the stellar masses of the supercluster galaxies;
iii) investigating the mechanisms driving galaxy evolution as function
of galaxy mass.  This work is carried out in the framework of the
joint research programme ACCESS aimed at determining the importance of
cluster assembly processes in driving the evolution of galaxies as a
function of galaxy mass and environment within the Shapley
supercluster (see Sect.~\ref{ACCESS}). In Sect.~\ref{sec:3}, we
describe the data-sets.  In Sect.~\ref{sec:4}, we derive the NIR
galaxy luminosity functions obtained through background subtraction in
the whole observed field and we study the ongoing effects of
environment by comparing the LFs of galaxies in three different
regions of the supercluster, characterized by high-, intermediate- and
low-density, we also compare NIR and optical LFs. The galaxy stellar
mass function is presented in Sect.~\ref{sec:5}.  The results are
discussed in Sect.~\ref{sec:6} and the summary and conclusions of this
work is given in Sect.~\ref{sec:7}.

Throughout the paper we adopt a cosmology with $\Omega_M$=0.3,
$\Omega_\Lambda$= 0.7, and H$_0$=70 km s$^{-1}$Mpc$^{-1}$. According to
this cosmology 1 arcmin corresponds to 60 kpc at $z$=0.048 and the
distance modulus is 36.66.

\section{The ACCESS project}
\label{ACCESS}
ACCESS\footnote{http://www.oacn.inaf.it/ACCESS} ({\it A Complete
Census of Star-formation and nuclear activity in the Shapley
supercluster}, PI: P. Merluzzi) is a project whose aim is to
distinguish among the mechanisms which drive galaxy evolution across
different ranges of mass examining how, when and where the properties
of galaxies are transformed by their interaction with the
environment. Since the most dramatic effects of environment on galaxy
evolution should occur in superclusters, where the infall and
encounter velocities of galaxies are greatest ($>$1000 km s$^{-1}$),
groups and clusters are still merging, and significant numbers of
galaxies will be encountering the dense intra-cluster medium (ICM) of
the cluster environment for the first time, we choose to study the
core region of the Shapley supercluster (Shapley
\citeyear{sha30}). The Shapley supercluster core is the richest and
most dynamically active region in the local Universe and hence
represents a unique laboratory for studying the effects of the
hierarchical assembly of structures on galaxy evolution.

The multi-wavelength data-set available for this project includes
panoramic imaging in UV [{\it Galaxy Evolution Explorer} (GALEX); PI:
R.J. Smith], optical [{\it European Southern Observatory} (ESO)
WideField Imager (WFI) archive data], near-infrared (UKIRT/WFCAM, PI:
R.J. Smith), mid-infrared ({\it Spitzer}, PI: C.P. Haines) all of which
cover at least a 2 deg$^2$ area of the SSC (see Fig. 1). Furthermore,
high-S/N medium-resolution optical spectroscopy (AAOmega, PI:
R.J. Smith) was obtained for a sample of 541 galaxies in the SSC field
(Smith, Lucey \& Hudson \citeyear{SLH07}; \citeyear{SLH09}), of which
448 are supercluster members ($0.039<z<0.056$). For this sample 371
galaxies have $B$, $R$ and $K$ photometry. The spectroscopic sample is
enlarged to $\sim$650 galaxies with published data, NASA Extragalactic
Database for bright objects, and 6dF data. New medium-resolution
integral-field spectroscopy will be provided by the Wide Field
Spectrograph (WiFeS, Dopita et al. \citeyear{DHM07}) at the Australian
National University 2.3\,m telescope.  A large programme of WiFeS
observations (PI: M. Dopita \& P. Merluzzi) started in April
2009. Finally, archive X-ray and radio data are also available. The
depth (e.g. $B$=22.5\,mag, $R$=22\,mag, $K$=18\,mag) and high-S/N of
the data allows the investigation of the photometric and spectroscopic
properties of supercluster galaxies well into the dwarf regime
(e.g. Mercurio et al. \citeyear{MMH06}; Gargiulo et
al. \citeyear{GHM09}).

The main scientific goals of ACCESS are: searching for the ram
pressure effects; probing galaxy merging and galaxy harassment and
``suffocation''; determining the frequency and the radial distribution
of cluster AGN; obtaining a statistical census of obscured
star-formation in cluster galaxies; correlate obscured star formation
with hierarchical cluster assembly; compare mid-IR, optical, radio and
UV star-formation indicators; investigating the fundamental plane of
low mass early-type galaxies.  Partners of this collaboration are the
Universities of Durham and Birmingham (UK), the Italian National
Institute of Astrophysics with the Observatory of Capodimonte and the
Australian National University (FP7-PEOPLE-IRSES-2008, Grant agreement
No. 230634).

\section{The data}
\label{sec:3}

The new near-infrared data analysed in this paper are complemented by
panoramic $B$- and $R$-band imaging from the Shapley Optical Survey
(SOS, Mercurio et al. \citeyear{MMH06} and Haines et
al. \citeyear{HMM06}, hereafter MMH06 and HMM06, respectively). In
Fig. \ref{fig1} the optical and NIR survey are superimposed. In the
following we describe the two surveys and their data products.

\subsection{The Shapley Optical Survey}
\label{sec:31}

The SOS comprises wide-field $B$- and $R$-band imaging covering a
2.0\,deg$^2$ region towards the clusters A3562, A3558 and A3556 which
form the core of the Shapley supercluster at z$\sim$ 0.05. 

The observations were carried out with the Wide Field Imager (WFI)
camera, a mosaic of eight 2046 $\times$ 4098 pixels CCDs giving a
field of view of 34$^\prime$ $\times$ 33$^\prime$, and mounted on the
Cassegrain focus of the 2.2m MPG/ESO telescope at La Silla. The survey
is made up of eight contiguous fields, each with total exposure times
of 1500\,s in $B$ band and 1200\,s in $R$ band, and typical FWHMs of
0.7-1.0\,arcsec. The data were retrieved from the ESO archive
and reduced using the {\sc ALAMBIC} pipeline (version 1.0, Vandame
\citeyear{van04}) and calibrated to the Johnson-Kron-Co\-u\-sins
photometric system using observations of Landolt (\citeyear{lan92})
standard stars. The sources are then extracted and classified using
SE{\sc xtractor} (Bertin \& Arnouts \citeyear{ber96}), resulting in galaxy
catalogues which are both complete and reliable (i.e. free of stars)
to $R$=22.0\,mag and $B$=22.5\,mag. Full description of the
observations, data reduction, and the production of the galaxy
catalogues are described in MMH06.

\subsection{The $K$-band survey}
\label{sec:32}
The $K$-band survey of the Shapley supercluster core was carried out
at the United Kingdom Infra-Red Telescope (UKIRT) with the Wide Field
Infrared Camera (WFCAM) in April 2007. The WFCAM instrument consists
of four 2048 $\times$ 2048 Rockwell detectors with a pixel scale of
0.4\,arcsec. The four detectors are spaced by 94\% of their active
area. A single exposure covers an equivalent area of 0.19\,deg$^2$ and
four interleaved exposures are required to achieve a filled tile of
0.865\,deg on a side (0.78\,deg$^2$). We observed a mosaic of five
complete tiles, covering a 3.043\,deg$^2$ (of which $\sim$2\,deg$^2$
overlap with the SOS) region centred on the SSC, which comprises three
Abell clusters A3556, A3558 and A3562 and two poor clusters SC
1327-312 and SC 1329-314, as shown in Fig. \ref{fig1}. The total
exposure time for each field is 300\,s, reaching $K$=19.5\,mag at
5$\sigma$, with typical FWHMs of 0.9-1.2\,arcsec.

The data were pipeline processed at WFCAM Science Archive
(WSA)/Cambridge Astronomy Survey Unit (CASU), reducing the frames and
performing astrometric and photometric calibration with respect to
2MASS (Irwin et al. \citeyear{irw04}). Zero-point uncertainty is
0.015\,mag and astrometry accuracy is $<$0.1\,arcsec (Irwin et
al. \citeyear{irw04}).

For each frame, a photometric catalogue was derived by using
SE{\sc xtractor} (Bertin \& Arnouts \citeyear{ber96}). We measured
magnitudes within a fixed aperture of 17\,arcsec diameter,
corresponding to $\sim $8\,kpc at $z{\sim}0.05$, and Kron (Kron
\citeyear{kro80}) magnitudes, for which we used an adaptive aperture
with diameter $a \cdot \mathrm{r}_{Kron}$, where $\mathrm{r}_{Kron}$
is the Kron radius and $a$ is a constant. We chose $a$ = 2.5, yielding
$\sim$ 94\% of the total source flux within the adaptive aperture
(Bertin \& Arnouts \citeyear{ber96}).  We measured the Kron magnitude
for all the objects in the catalogue and adopted it as the total
magnitude. Luminosity functions were computed by means of Kron
magnitudes, while aperture magnitudes were used for measuring galaxy
colours. Since we derive galaxy colours using the same apertures both
at optical and NIR wavelengths, we checked the effects of seeing
variations among the wavebands by degrading the $R$-band image to the
seeing of the $K$-band image. Comparing the aperture magnitudes in the
original and the degraded image we find a difference which is an order
of magnitude lower than the photometric error, as expected since the
aperture is large compared to the seeing.

Particular care is needed to avoid stellar contamination due to the
high number density of both stars and galaxies in this field (the
Galactic latitude of this field is +30$^\circ$) which increases the
frequency of star-star and star-galaxy blends that can be
misclassified as a single galaxy. For the star/galaxy classification
we make use of the optical photometry when available in the sense that
objects observed in both $R$ and $K$ bands were classified as stars
and galaxies according to the $R$-band classification\footnote{The
$R$-band WFI mosaic is the deepest and highest resolution
data obtained for this project, reaching M$^\star$+7, and is therefore
generally used as reference for our multi-wavelengths surveys.}. As
shown in Tab.~3 of MMH06, at $R=20.0-20.5$\,mag, only 2\% of the stars
were misclassified as galaxies. According to the typical $R-K$ colour
of early-type galaxies at $z{\sim}0.05$ (Poggianti \citeyear{pog97}),
this magnitude range corresponds to $K\sim 17.5-18.0$\,mag. The
contamination of misclassified stars is taken into account in the
galaxy LF determination. For those objects observed in the $K$-band
having no optical magnitude measurement, we use the distribution of sources in
the stellarity index (SI) parameter of SE{\sc xtractor} versus Kron
magnitude ($K$) diagram to separate stars and galaxies. We classified
as stars those objects whose SI value is larger than a given
threshold: SI$_{\mathrm{min}}$=0.8. This value of SI$_{\mathrm{min}}$
has been chosen by adding simulated stars and galaxies to the $K$-band
images, and measuring their SI and $K$ parameters by means of
SE{\sc xtractor}, in the same way as for real sources. Simulated stars and
galaxies were randomly generated in a magnitude range of
$K$=12-20\,mag using the software 2DPHOT (La Barbera et
al. \citeyear{lab08}).

The completeness of the $K$-band catalogue was estimated by measuring
the percentage of simulated galaxies and stars which are recovered by
SE{\sc xtractor} as a function of $K$-band magnitude. The
completeness function was found to be strongly dependent on the source
density and therefore is different for each $K$-band frame. In fact,
in high density regions the catalogue is $\sim$ 65\% and $\sim$ 50\%
complete at $K$=17.5\,mag and $K$=18.0\,mag, respectively, while in
low density regions it is $\sim$ 100\% and $\sim$ 80\% complete at
$K$=17.5\,mag and $K$=18.0\,mag, respectively. We correct galaxy
counts using a different completeness function for each $K$-band
frame, i.e. by taking both the crowdedness and magnitude effects fully
into account for each galaxy, when determining its contribution to the
counts in a given magnitude bin. By weighting each galaxy by the
locally estimated completeness of the survey, we are able to obtain
unbiased comparisons of the galaxy counts in different environments.
Since in the high density region the completeness is less than 50\%
beyond $K$=18.0\,mag, we adopted this conservative limit as the
magnitude to which catalogues can be reliably corrected for
incompleteness when determining the galaxy LF. At this limiting
magnitude the accuracy of the completeness functions is better by 5\%
and is taken into account in the error budget of each galaxies. The
final catalogues consist of 18,534 galaxies with $K\leq$18\,mag.

\section{NIR Luminosity Functions}
\label{sec:4}

The $K$-band galaxy LF of the SSC has been derived down to the
magnitude limit $K$=18\,mag accounting for interlopers by the
statistical subtraction of the background contamination. We chose this
approach since we do not have spectroscopic information
complete for whole the galaxy sample in the considered magnitude
range (being $\sim$90\% complete for $R<$16\,mag or $K<$13.5) and the available
photometry ($B$, $R$ and $K$ band) does not allow us to derive
accurate photometric redshifts. The number counts in the supercluster
field has been obtained by weighting each galaxy's contribution to a
given magnitude bin according to its completeness.  We also correct
for the contamination of the misclassified stars.  Absolute magnitudes
were determined using the k- and evolutionary corrections for
early-type galaxies at $z\sim$0.05 from Poggianti (\citeyear{pog97}).
The large observed area and the depth of the survey are suitable to
investigate the effects of environment within the supercluster.

\subsection{Background galaxy subtraction}
\label{sec:41}
Since the area covered by the SSC observations lies completely within
the overdensity corresponding to the core complex, it is not possible
to use the outer regions of the survey to estimate the
background/foreground contribution to the galaxy counts. Therefore, we
performed the statistical subtraction of field galaxies by means of a
control field observed with the same instrument at a suitable depth. A
similar or larger area is necessary in order to reduce the effects
of field-to-field variance and small number statistics.

   \begin{table}
     \caption[]{Central coordinates of DXS fields used for the
       statistical background subtraction.}

     $$
           \begin{array}{c c c}
            \hline
            \noalign{\smallskip}
            \mathrm{Name} &   \mathrm{RA (J2000)} & \mathrm{DEC (J2000)}\\
                \noalign{\smallskip}
                \hline
                \hline
            \noalign{\smallskip}

\mathrm{XMM-LSS}        & 02 \ \ 25 \ \ 00 & -04 \ \ 30 \ \ 00\\
\mathrm{ELAIS N1}        & 16 \ \ 10 \ \ 00 & +54 \ \ 00 \ \ 00\\
\mathrm{SA22}                & 22 \ \ 17 \ \ 00 & +00 \ \ 20 \ \ 00\\

            \noalign{\smallskip}
            \hline
            \noalign{\smallskip}
            \hline
         \end{array}
    $$
      \label{tab1}

   \end{table}

\begin{figure}
\centerline{{\resizebox{8cm}{!}{\includegraphics{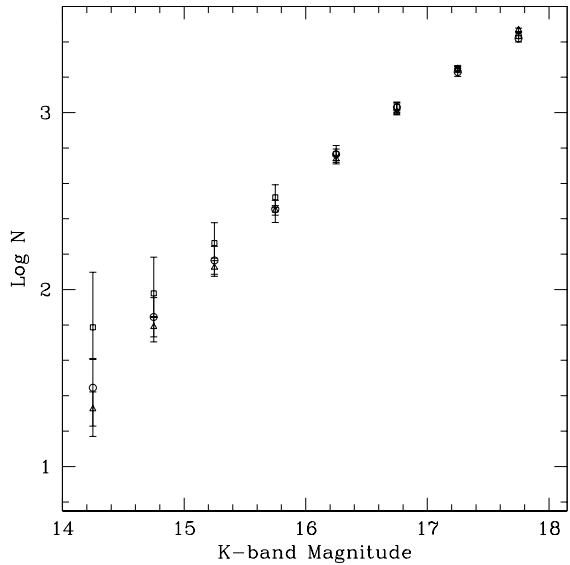}}}}
\caption{Comparison of galaxy counts obtained from the DXS (open
triangles), CADIS (open circles) and ALHAMBRA (open squares). The counts are
normalized to the total area covered by the CADIS data.}
\label{fig2}
\end{figure}

To this aim we chose the UKIRT Infrared Deep Sky Survey (UKIDSS) Deep
Extragalactic Survey (DXS) (Lawrence et al. \citeyear{law07}) which
aims to map 35\,deg$^2$ of sky to a magnitude limit of $K$=20.8\,mag
at 5$\sigma$. Since our $K$-band photometry is 50\% complete down to
$K$=18.0\,mag, the UKIDSS DXS data are suitable to estimate
field galaxy counts.

The background contamination was estimated from three control fields
from the UKIDSS DXS 3rd data release (Warren et al., in preparation)
reaching the required depth. In particular, we considered 24
multiframes of $\sim$ 0.19\,deg$^2$ , covering a total area of
$\sim$4.55\,deg$^2$ over three regions of sky (Tab.~\ref{tab1}), with
exposure times of $\gtrsim$ 360\,s in $K$. We note that this area is
significantly larger than that covered by any published table of
galaxy counts reaching $K=18$ or deeper, the surveys of V\"ais\"anen
et al. (\citeyear{VTW00}) and K\"ummel \& Wagner (\citeyear{KW01})
limited to areas of $\sim0.9$\,deg$^{2}$, and reaching just
$K=$17-17.5. The $K$-band catalogues were obtained from the WFCAM
Science Archive (WSA, Hambly et al. \citeyear{ham08}) selecting from
the dxsDetection table the isolated (pperrbits$\le$1) objects with
non-stellar morphologies (class=1).

Since we have accounted statistically for the incompleteness affects
  in the $K$-band survey (see Sect.~\ref{sec:32}), we can subtract the
  control field counts from those obtained in the Shapley area in
  order to obtain supercluster member counts.  In this case, following
  Bernstein et al. (\citeyear{ber95}) the background counts are
  estimated as the mean of the control field counts corrected for the
  ratio between the observed areas (Eq. 1 of Bernstein et
  al. \citeyear{ber95}) and errors on the background counts are
  estimated through an empirical approach as the {\it rms} of the
  counts in each control field with respect to the mean estimated over
  the whole area (Eq. 2 of Bernstein et al. \citeyear{ber95}).  For
  the background counts, scaling the {\it rms} value to the whole
  area, we obtain an uncertainty on the mean counts of $\sim$11\% at
  $K<$ 13.5\, mag where small number statistics dominates, and less
  than 3\% at $K$=18\,mag. For a Gaussian distribution, the standard
  deviation estimated from three samples is within a factor of two
  from the true value, in 80\% of trials. Since each field is smaller
  than the $K$-band survey, we overestimate the field-to-field
  variance among larger fields, taking into account that the
  fluctuations due to galaxy clustering are smaller for wider
  fields. Following Ellis \& Bland-Hawthorn (\citeyear{EBH07}) we
  estimate that the field-to-field variance is overestimated of about
  the 15\%. Using this method the error on background counts accounts
  for the Poissonian fluctuations and the field-to-field variance. We
  also include the error related to the photometric uncertainties
  which is estimated by means of 100,000 Monte-Carlo simulations,
  changing the galaxy luminosities according to the photometric error
  and re-computing the number counts in each magnitude bin.

These galaxy number counts were found to be consistent with field
galaxy counts from the Calar Alto Deep Imaging Survey (CADIS, see
Huang et al. \citeyear{HGC03}), a medium deep $K$-band survey with a
total area of 0.2 deg$^2$ and a completeness magnitude of 19.75\,mag
(see Fig.~\ref{fig2}). We also agree with the counts of the Advanced
Large Homogeneous Area Medium-Band Redshift Astronomical (ALHAMBRA)
survey given by Crist\'obal-Hornillos et al. (\citeyear{cri09})
covering a 0.44\,deg$^2$ and reaching K$_S$=19.5\,mag. We note that
Ellis \& Bland-Hawthorn (\citeyear{EBH07}) who have combined numerous
published galaxy number counts over a wide range of pass-bands and
magnitudes, obtain galaxy counts in the range $13<K<18$ that are
10-15\% higher than ours, but are based on small patches of sky
($<1$\,deg$^{2}$) and show significant variations from one magnitude
bin to another as surveys fall out of the sample.

The counts of SSC galaxies were defined as the difference between the
counts detected in the supercluster fields and those estimated for the
background (Eq. 3 of Bernstein et al. \citeyear{ber95}). Then, the
uncertainties were measured as the sum in quadrature of fluctuations in
the background and in the supercluster counts (Eq. 4 of Bernstein et
al. \citeyear{ber95}).

\begin{figure*} 
\centerline{\resizebox{10cm}{!}{\includegraphics[angle=-90]{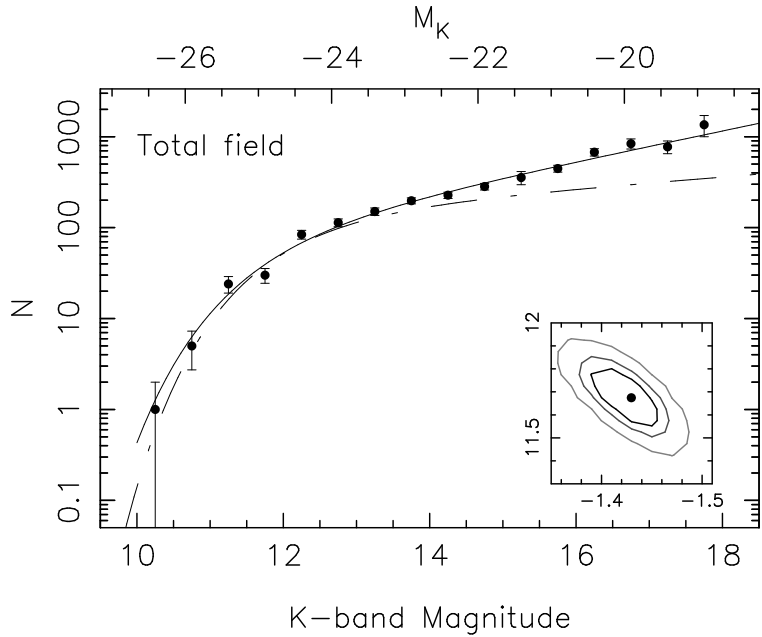}}}
\caption{$K$-band LF of the whole SSC and its fit with the Schechter
function, continuous line. The counts are per half magnitudes bins.
The dot-dashed lines is the fit to the $K$-band LF in the field of
Jones et al. \citeyear{JPC06} normalized to the luminosity density
$\phi$ of SSC LF. In the small panel the 1, 2 and 3$\sigma$ confidence
levels of the best-fit parameters for $\alpha$ (x-axis) and m$^*$
(y-axis) from the Schechter fit, are shown.}
\label{totLF} 
\end{figure*}

\subsection{The total luminosity function}
\label{sec:42}

Figure~\ref{totLF} shows the Shapley LF over the whole $K$-band survey
covering the SSC (Fig.~\ref{fig1}), and the fit with a single
Schechter function. The error bars on number counts take into
account Poissonian uncertainties, field-to-field variance (see
Sect.~\ref{sec:41}), photometric uncertainties and, in the last bin,
the percentage of misclassified stars (see Sect.~\ref{sec:32}).

The LF fit was obtained with a $\chi^{2}$ minimization routine,
accounting for the finite size of magnitude bins by integrating the
Schechter function over each magnitude bin. The Schechter function fit, whose
parameters are the faint-end slope, $\alpha$, the characteristic
magnitude, M$^{*}$, and the luminosity density $\phi$, according to
the $\chi^{2}$ statistics, provides an acceptable description of
cluster galaxy counts. The best-fitting values of the $K$-band Schechter
function parameters are $M_{K}^{*}=-24.96\pm0.10$ and
$\alpha=-1.42\pm0.03$.  The confidence contours on $\alpha$ and m$^*$
are shown in the small panel in Fig.~\ref{totLF} and were derived by
randomly shifting galaxy number counts according to their
uncertainties, and re-computing the best-fitting Schechter function for each
realization. The fit parameters and associated $\chi^{2}$ statistics
are listed in Tab.~\ref{fitsLF}.

The NIR LF trend derived in the 3\,deg$^2$ area of the SSC agrees
with the results of Mobasher $\&$ Trentham (\citeyear{MT98}) who found
$\alpha$=--1.41 in the magnitude range --19.5$<M_K<$--16.5, although
they studied only a 41.1\,arcmin$^2$ region of the Coma cluster core
corresponding to about 10\,arcmin$^2$ at Shapley redshift.

{In Fig.~\ref{totLF} we plot for comparison the field $K$-band LF
by Jones et al. (\citeyear{JPC06}) who found M$^\star_K=-24.60$ and
$\alpha=-1.16$. The two LFs appear consistent down to $M^{*}_{K}=-23$,
but the faint-end slope for the SSC is steeper than that measured in
the field at $>3\sigma$ level. The field value of $M_{K}^{*}$ is also
not consistent with that obtained for the SSC with a single Schecter
function, with $M^{*}_{K}$ being brighter for the SSC than for the
field, at the $3\sigma$ level according to the confidence contours in
Fig. 10 of Jones et al. (\citeyear{JPC06}).

   \begin{table*}
     \caption[]{Fits to the LFs. Errors on the $\mathrm{m^*}$ and
      $\alpha$ parameters can be obtained from the confidence contours
      shown in Figs.~\ref{totLF} and \ref{contK}. Column 1: the
      analyzed region. Column 2: the fitting function. S for a single
      Schechter, G+S for Gaussian plus Schechter and S+S for two
      Schechter functions. Columns 3 and 4: characteristic magnitude
      of the Schechter function (apparent m$^\star$ and absolute
      M$^\star$) or central magnitude $\mu$ of the Gaussian
      function. Column 5: the value of slope $\alpha$ or amplitude
      $\sigma$ of the Schechter and the Gaussian function,
      respectively. Columns 6, 7 and 8: the same of columns 3, 4 and 5
      for the faint-end of the LF. Column 9 and 10: reduced $\chi^2$
      and its probability.}

     $$
           \begin{array}{c c | c c c | c c c | c c }
            \hline
            \noalign{\smallskip}
            \mathrm{Region} & \mathrm{Function} &
            \mathrm{m^*}/\mu & \mathrm{M^*}/\mu &\alpha/\sigma & 
\mathrm{m^*_f}   &  \mathrm{M^*_f} &\alpha_f &
\mathrm{\chi^2_{\nu}} & \mathrm{P(\chi^2>\chi^2_{\nu})}\\
                \noalign{\smallskip}
                \hline
                \hline
            \noalign{\smallskip}
            \mathrm{all \ \ field}  & \mathrm{S}   & 11.70 & -24.96 &
-1.42 &        &         &      & 1.29& 21.21\% \\
                                    && &&&&&&&\\
            \mathrm{high \ \ density}  & \mathrm{S} & 11.48 & -25.18 &
-1.33 & & & & 0.77& 69.31\% \\
            \mathrm{high \ \ density}  & \mathrm{G+S} & 13.81 & -22.85  & 
 \ \ 1.97  & 12.58\pm0.12 & -24.08 & -1.65\pm0.03 & 0.58 &
83.46\% \\
            \mathrm{high \ \ density}  & \mathrm{S+S} & 11.57 & -25.09  &
-1.22  & 16.64\pm0.29 & -20.02 & -2.05\pm0.50  & 0.74 &
68.28\% \\
                                    && &&&&&&&\\
            \mathrm{int \ \ density}  & \mathrm{S} & 11.47   & -25.19 &
-1.44  & & & & 1.30 & 20.39\% \\
            \mathrm{int \ \ density}  & \mathrm{G+S} & 13.81 & -22.85  &  
\ \ 1.90 & 12.47\pm0.11 & -24.19  & -1.64\pm0.02 & 1.44 &
15.31\% \\
            \mathrm{int \ \ density}  & \mathrm{S+S} & 11.98 & -24.68  &
-1.26 & 14.44\pm0.21 & -22.22  & -2.41\pm0.11 & 1.34 & 20.24\%
\\
                                    && &&&&&&&\\
            \mathrm{low \ \ density}  & \mathrm{S} & 11.51   & -25.15  &
-1.49   & & & & 1.00 & 44.78\% \\
            \mathrm{low \ \ density}  & \mathrm{G+S} & 13.79 & -22.87  &  
\ \ 1.58 & 14.13\pm0.11 & -22.52 & -1.74\pm0.04 & 0.20 &
99.60\% \\
            \mathrm{low \ \ density}  & \mathrm{S+S} & 12.50 & -24.16  &
-1.04 & 15.72\pm0.14 & -20.94 & -1.90\pm0.12 & 0.18 & 99.77\%
\\
            \noalign{\smallskip}
            \hline
            \noalign{\smallskip}
            \hline
         \end{array}
    $$
      \label{fitsLF}

   \end{table*}

\begin{figure*}
\centerline{{\resizebox{\hsize}{!}{\includegraphics[angle=-90]{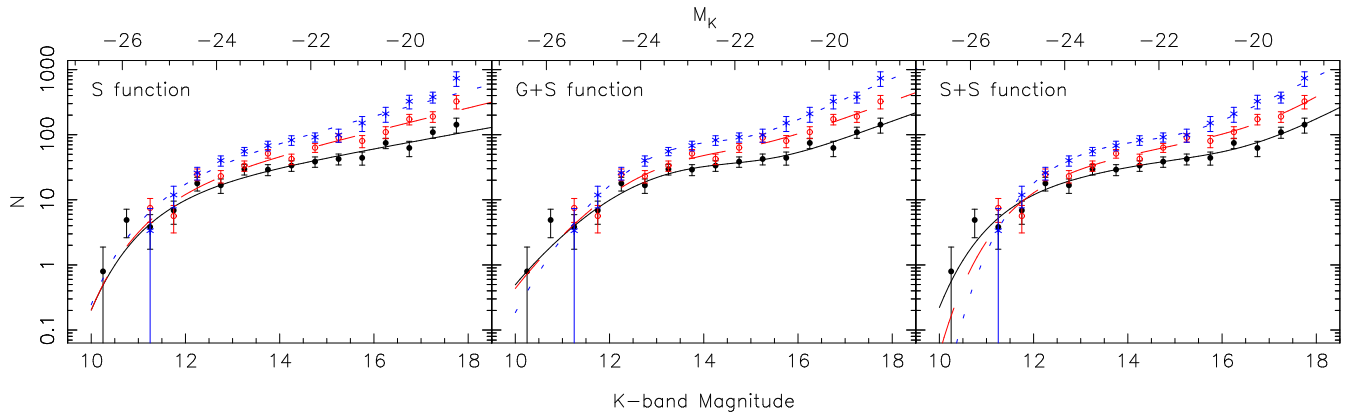}}}}
\caption{
The $K$-band LFs of galaxies in the three cluster regions
corresponding to high- (filled circles), intermediate- (open circles)
and low-density (crosses) environments. In the left, central and right
panels the continuous, long-dashed and short-dashed lines represent
the fits with Schechter, G+S and S+S functions to high-, intermediate-
and low-density regions, respectively. The counts are per half
magnitude bin.}
\label{LFenv}
\end{figure*}

\begin{figure}
\centerline{{\resizebox{8cm}{!}{\includegraphics[angle=-90]{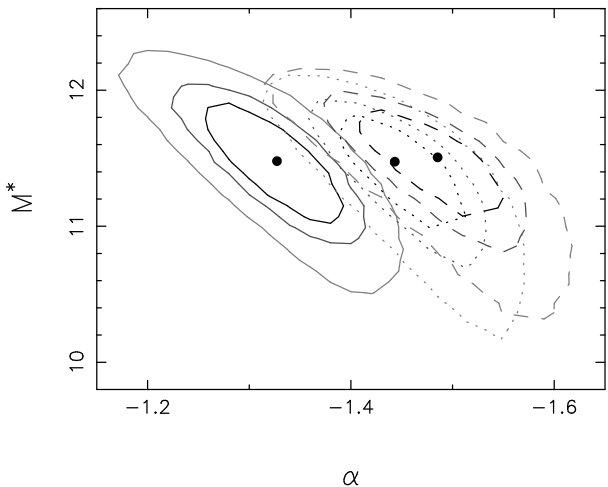}}}}
\caption{The 1, 2 and $3\sigma$ confidence regions for the $K$-band
Schechter parameters (right panel in Fig.~\ref{LFenv}) for the three
cluster regions corresponding to high- (solid contours), intermediate-
(dotted) and low-density (dashed) environments.}
\label{contK}
\end{figure}

\subsection{The effect of environment}
\label{sec:43}

Balogh et al. (\citeyear{BCZ01}), using 2MASS data, were the first to
detect environmental variances in the NIR LF. This environmental
effect was then questioned by Rines et al. (\citeyear{RGD04}) who
found that the cluster LFs derived in the virial regions and in the
infall regions are very similar, although both poorly fitted by a
Schechter function. However, in their work, which is also based on the
2MASS survey, they noted that at magnitudes fainter than the
completeness limit the LF in the infall regions may indicate a steeper
faint-end slope which should be investigated by means of deeper
data-sets.

In order to investigate the effects of the environment we derived and
compared the LFs in three different regions of the supercluster,
characterised by high-, intermediate- and low-densities of galaxies
(see Fig.~\ref{fig1}) where galaxy densities are $\rho>$1.5,
1.0$<\rho\leq$1.5 and 0.5$<\rho\leq$1.0 gals arcmin$^{-2}$,
respectively. The local density of $R<21$\,mag galaxies, $\Sigma$, was
determined across the $R$-band WFI mosaic (i.e. in a 2 deg$^2$ area,
see Fig. 1). We derive $\Sigma$ by using an adaptive kernel estimator
(Pisani \citeyear{pis93}; \citeyear{pis96}), in which each galaxy $i$
is represented by a Gaussian kernel,
$K(r_i)\propto\exp(-r^{2}/2\sigma_{i}^{2})$, whose width $\sigma_{i}$
is proportional to $\Sigma_{i}^{-1/2}$ thus matching the resolution
locally to the density (see MMH06 for more details).

In left-hand panel in Fig.~\ref{LFenv}, left panel, we show the
$K$-band LFs of galaxies in the high- (filled circles), intermediate-
(open circles) and low-density (crosses) regions covering areas (in
the SOS/$K$-band survey overlap) of $\sim$0.115, 0.330 and
1.062\,deg$^2$, respectively, together with their fits with a
Schechter function (continuous, long-dashed and short-dashed lines,
respectively). The background subtraction was performed as for the
total LF simply scaling the counts with the area values because of the
complex geometry of the three density regions. Figure~\ref{contK}
shows the confidence contours of the best fitting Schechter function
for the three density regions. The faint-end slope becomes steeper
from high- to low-density environments varying from --1.33 to --1.49,
being inconsistent at the 2$\sigma$ confidence level (c.l.) between
high- and low-density regions. We note that in this investigation the
results are also more robust than those obtained for the total LF
against variations in background galaxy counts, since we are
considering denser supercluster regions and, furthermore the control
field is significantly larger than the regions of different
environments.

This result seems to be in contradiction with previous works (e.g. Balogh et
al. \citeyear{BCZ01}; Croton et al. \citeyear{CFN05}) which found a LF
trend that varies smoothly with local density and/or environment. It
should be note that these studies are based on the 2MASS survey
which allows to investigate the luminosity distribution only down to
about M$^\star$+2 at the redshift of Shapley, while we are considering
a much fainter galaxy population which is more likely affected by
environmental related processes (e.g. Haines et al. \citeyear{HGL07}).

It is also worth pointing out that comparing the flat NIR LF measured
for field galaxies by recent surveys (e.g. 6dF Galaxy Survey, Jones et
al. \citeyear{JPC06}) with our slope in the low-density environment,
one has to take into account that, since we are still in the cluster
environment, this region can be suitably associated to the infall
region rather than to the field.  On the other hand, the LF slopes
obtained by CAIRNS (Cluster And Infall Region Nearby Survey; Rines et
al. \citeyear{RGD04}) in both cluster virial and infall regions are
consistent with the value derived for the overall SSC with a single
Schechter function.

\begin{figure*}
\centerline{\resizebox{\hsize}{!}{\includegraphics[angle=-90]{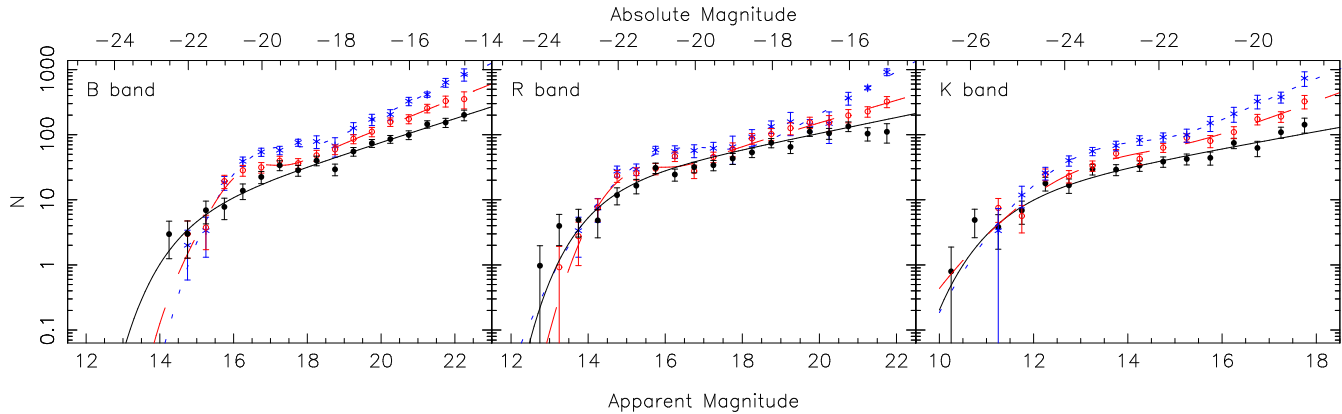}}}
\caption{Galaxies LFs $B$- (left panel), $R$-band (central panel) and
$K$-band (right panel) high- (filled circles), intermediate- (open
circles) and low-density (crosses) environments. The long-dashed and
short-dashed lines represent the the G+S best fit for intermediate-
and low-density environments respectively and the continuous line is
the Schechter best fit for high-density regions.}
\label{LFconfr}
\end{figure*}

According to the $\chi^2$ statistics, in all the three environments
the fit with a single Schechter function cannot be rejected, but there
is some ``structure" evident in the residuals: the fit systematically
under- and over-predicts the observed counts as a function of
magnitude. The LFs suggest instead a bimodal behaviour due to the
presence of an upturn for faint galaxies, that cannot be described by
using a single Schechter function. To successfully model these changes
in slope and to compare our results with our optical LFs (see
Sect.~\ref{sec:44}), we fit our data with a composite Gaussian +
Schechter (G+S) LF (Fig.~\ref{LFenv}, central panel) and the sum of
two Schechter (S+S) functions (Fig.~\ref{LFenv}, right panel). Looking
at Fig.~\ref{LFenv} and Tab.~\ref{fitsLF} we note that the G+S and
S+S fits significantly improve the data description, particularly in
the low-density region, where according the reduced $\chi^2$ the
probability of the fit is P($\chi^2>\chi^2_\nu$) $\sim$ 99.6\% and
99.8\% for the G+S and S+S functions against P($\chi^2>\chi^2_\nu$)
$\sim$ 46\% for the single Schechter function. Since the function
fitting the faint-end is poorly constrained by the data (the faint
component dominates only in the last three magnitude bins), in order
to estimate the uncertainties on the parameters characterizing the LF
at fainter luminosities, m$^\star_{\mathrm{f}}$ and
$\alpha_{\mathrm{f}}$, we proceed as follows. Fixing all the best fit
parameters except m$^\star_{\mathrm{f}}$ or $\alpha_{\mathrm{f}}$, we
randomly shift galaxy number counts according to their uncertainties,
and then re-compute the best-fitting functions obtaining a range of
values for m$^\star_{\mathrm{f}}$ and $\alpha_{\mathrm{f}}$. The
faint-end slope becomes steeper from high-/intermediate- to
low-density environments varying from --1.65 to --1.74 with the G+S
fit.

\subsection{Comparison with Optical LFs}
\label{sec:44}

In order to investigate the processes responsible for changing
the shape of the galaxy LF, we compare the trends observed for the
optical and the NIR LFs. Since the NIR LF is expected to
approximate the SMF, while the optical LFs are more sensitive to the
galaxy star-formation history, both are needed for investigating the
nature of galaxies which dominate the faint-end. For instance, a steep
optical LF can be compatible with a flat stellar mass function if dwarf
galaxies have their luminosities boosted by starbursts, a scenario 
which can be probed by comparison to the NIR LF.

The optical LF of the SSC derived in the SOS (see MMH06) cannot be
described by a single Schechter function due to the dips apparent at
M$^\star$+2 both in $B$ and $R$ bands and the clear upturn in the
counts for galaxies fainter than $B$ and $R\sim$18\,mag. Instead the
sum of a Gaussian and a Schechter function, for bright and faint
galaxies, respectively, is a suitable representation of the data.
Furthermore, we observed significant environmental trends in the form
of a dip which becomes deeper, and a faint-end slope which becomes
steeper, with decreasing density. In particular, the slope values
becomes significantly steeper from high- to low-density environments
varying from --1.46 to --1.66 in $B$ band and from --1.30 to --1.80 in
$R$ band, being inconsistent at more than 3$\sigma$ c.l. in both
bands. Such a marked luminosity segregation is related to the
behaviour of the red galaxy population: while red sequence counts are
very similar to those obtained for the global galaxy population, the
blue galaxy LFs are well described by single Schechter functions and
do not vary with the density. We explained these results in terms of
the galaxy harassment scenario, in which the late-type spirals that
represent the dominant population at $\sim$M$^\star$+2 are transformed
by galaxy harassment into passively-evolving dwarf spheroids, and in
the process become $\sim$1-2 magnitudes fainter due to mass loss and
an ageing stellar population without new star formation. The
observed changes in the shape of the LF can be considered as
reflecting the changes in the mixture of galaxy morphological types
with environment described by the morphology-density relation, from
late-type dominant (and hence a steep faint-end to the LF) in
low-density regions to early-type dominant (with a shallower faint-end
to the LF), in the cluster cores (Binggeli, Sandage \& Tammann
\citeyear{BST88}; de Lapparent \citeyear{deL03}).

In Fig.~\ref{LFconfr} we compare the shape of the composite G+S $B$
(left panel), $R$ (central panel) LFs, derived in MMH06, with the
relative $K$-band LFs (right panel) in the three density regions. Both
LFs (optical and NIR) are steeper in the low-density regions.
However, the NIR LFs only marginally indicate an absence of
M$^\star$+2 galaxies in low-density environment.

The study of the optical LFs showed that it is the red galaxy
populations that turn out to contribute to the change of the LF
trend with environment, in particular to the steepening of the
faint-end slope. This is at odds with what is found in the field where
the faint-end is populated by galaxies of smaller masses, later
morphologies, bluer colours, later spectral types, and stronger line
emission (e.g. Balogh et al. \citeyear{BCZ01}; Madgwick et
al. \citeyear{mad02}). On the other hand, in clusters the LF of
early-type galaxies is found to be steeper than in the field (De
Propris et al. \citeyear{DCD03}).  We should also take into account
that at least part of the red galaxies contributing the Shapley LFs in
different supercluster environments may not be passive: Haines et
al. (\citeyear{HGM08}) studying a volume-limited sample
(0.005$<z<$0.037) of local galaxies found that $\sim$30$\%$ of red
sequence galaxies in the optical colour-magnitude diagram show signs
of ongoing star-formation from their spectra and this contamination is
greater at faint magnitudes (M$_r >$--19\,mag).

\section{The galaxy stellar mass function}
\label{sec:5}

The combined optical and NIR data allow us to derive the distribution
of galaxy stellar masses. The sample we analysed is in the magnitude
range 10$\leq K \leq$18 and refers to the $\sim$2 deg$^2$ area covered by both
the SOS and our $K$-band imaging.  We note that although the area
of the NIR survey is slightly different from that of the SOS, both
surveys map the same kinds of environment from the low- to the
high-density (see the isodensity contours in Fig.~\ref{fig1}).

One of the main concerns in deriving the stellar mass function of SSC
galaxies is the foreground/background contamination that has to be
estimated and corrected for. There exists already a wealth of
spectroscopic data in the region covered by optical and NIR
photometry, comprising about 650 galaxy redshifts (see
Sect.~\ref{ACCESS}) corresponding to $\sim$ 90 \% of $R<$ 16\,mag
galaxies. In order to extend the magnitude range and improve the
statistics, we need to adopt a complementary approach for those
galaxies without spectroscopic information. We use the probability
that galaxies are supercluster members as derived by HMM06 following
Kodama \& Bower (\citeyear{KB01}).  We consider separately the three
cluster environments as well as the remaining galaxies in the SOS, and
construct two-dimensional histograms with bins of width 0.4\,mag in $R$
and 0.2\,mag in $B-R$ to properly map the galaxy colour-magnitude
distribution. The number counts in each bin are then compared with
those expected for a suitable field region, normalized to have the
same overall area. For this purpose we used a 4.3\,deg$^{2}$ region of
deep $BVR$ imaging from the Deep Lens Survey (Wittman et
al. \citeyear{wit02}).  The probability that a randomly selected SOS
galaxy belongs to the supercluster is then determined from the ratio
of the number counts obtained for that bin from the SOS and DLS (Eq. 1
of HMM06). For those galaxies with available redshifts the probability
is set to 1 for $0.035{<}z{<}0.056$\footnote{The redshift range is
derived from the redshift distribution of galaxies in the SOS field
with available spectroscopy.} or 0 otherwise.

The stellar masses of galaxies belonging to the Shapley supercluster,
according to the previous criteria, contribute to the galaxy stellar
mass function according to their likelihood of belonging to the
Shapley supercluster.

\begin{figure}
\centerline{\resizebox{10cm}{!}{\includegraphics[angle=-90]{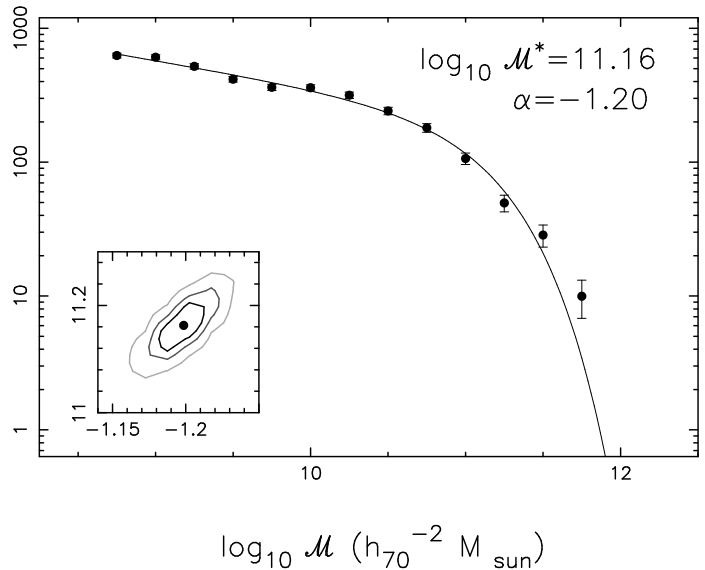}}}
\caption{Stellar Mass Function of galaxies in the  SSC (see
text).The continuous line indicates the best single Schechter fit to
the data. In the small panel the 1, 2
and 3$\sigma$ c.l.s of the best-fitting parameters for $\alpha$
(x-axis) and log$_{10}$ $\mathcal{M}^\star$ (y-axis) are shown.}
\label{GSMF}
\end{figure}

\subsection{Derivation of stellar masses}
\label{sec:51} 
 The stellar masses of galaxies belonging to the Shapley supercluster
are estimated by means of stellar population models constrained by the
observed optical and infrared colours. It is well known that stellar
masses estimated using the fit to the multicolour spectral energy
distribution (SED) are model dependent and subject to various
degeneracies. In order to reduce such degeneracies we have used a
large grid of complex stellar population models by Maraston
(\citeyear{mar05}) with a Salpeter initial mass function covering a
wide range of parameters. We use SEDs with the star-formation history
(SFH) parameterised as $\Psi(t) \propto \mathrm{exp} (-t/\tau)$, with
$\tau$ between 1.0 and 20.0 Gyr, ages between 0.001 and 14 Gyr and
metallicities in the range 0.5-2.0 Z$_{\odot}$. We note that the
metallicity of low-mass galaxies can be lower (e.g. Smith et
al. \citeyear{smi09}), but at present complex stellar population
models do not explore such a low metallicity range. The synthetic
spectra are shifted to the galaxy spectroscopic redshift, if known, or
at the median supercluster redshift $z$=0.05. Then for each of them we
compute the $B$, $R$ and $K$ magnitudes by adopting the Calzetti
(\citeyear{cal01}) extinction. Since in the photometric calibration of
the $B$ band the colour term ($B-R$) is not negligible, as discussed
in MMH06, we use the WFI $B$ filter in order to compute $B$
magnitudes\footnote{Our B-band photometry agrees within the zero-point
uncertainties ($\sim$0.03 mag) with that obtained by the WIde-field Nearby
Galaxy cluster Survey (WINGS; Varela et
al. \citeyear{var09}) using independent WFI B-band observations for a
subset of the SOS field.} from the models.

\begin{figure*}
\centerline{{\resizebox{\hsize}{!}{\includegraphics[angle=-90]{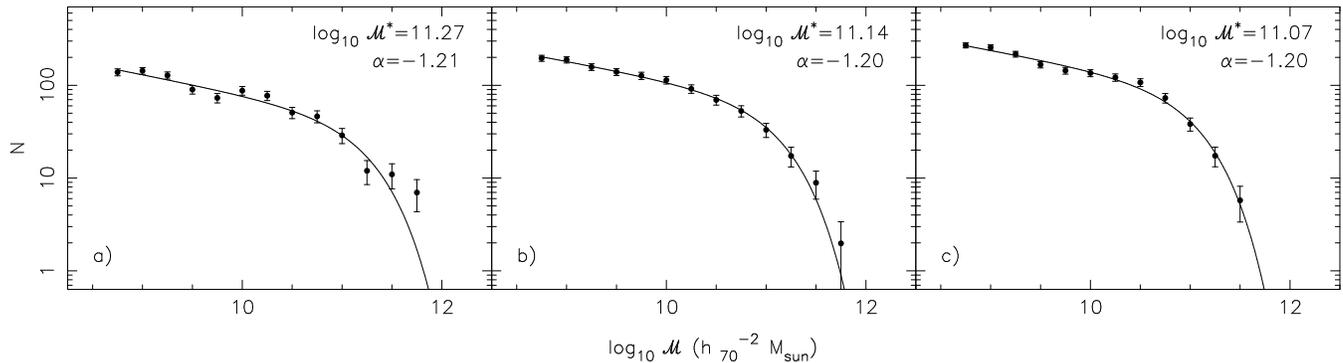}}}}
\caption{The mass function of galaxies in the three cluster regions
corresponding to high- (panel a) , intermediate- (panel b) and low-density
(panel c) environments. In the left, central and right panel the
continuous line represents the fit to the data. In each panel the best fit
value of $\alpha$ and log$_{10}$ M$^*$ are reported.}
\label{MFenv}
\end{figure*}

\begin{figure*}
\centerline{{\resizebox{8cm}{!}{\includegraphics[angle=-90]{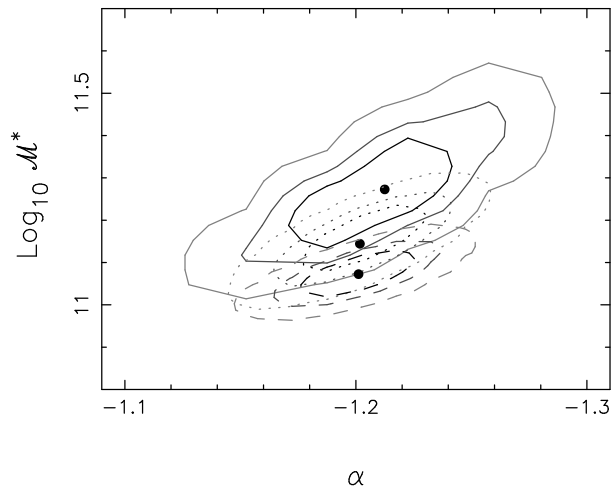}}}}
\caption{The 1, 2 and $3\sigma$ confidence levels for the Galaxy
Stellar Mass Function parameters in the three cluster regions
corresponding to high- (solid contours), intermediate- (dotted) and
low-density (dashed) environments.}
\label{MFcont}
\end{figure*}

The most appropriate evolutionary history is selected by fitting the
optical + NIR photometry, and the mass is estimated by normalising the
best-fit SED to the observed $K$-band magnitude.  This choice of model
grid parameters yields a fairly uniform coverage of colour space and
well represents the SEDs of the galaxies in the sample.  The
uncertainty on the resultant stellar mass was estimated by performing
Monte Carlo simulations, shifting the galaxy colours according to
their corresponding uncertainties, and re-computing each time the
mass. Further sources of uncertainty in the fitting technique are the
error on redshift, which is fixed at $z$=0.05 for those galaxies
without available spectroscopic redshift. We perform Monte-Carlo
simulations by randomly shifting SEDs in the redshift range
0.035$<z<$0.056 adopted to select spectroscopic confirmed supercluster
members. This contribution to the mass uncertainty is negligible,
since it is at least one order on magnitude lower than that due to the
photometric errors. Besides of this, the main source of uncertainty is
the adopted IMF (see Bell \& de Jong \citeyear{BdJ01}). Bell et
al. (\citeyear{BMK03}) found that, using the simple stellar populations
with different IMF, stellar masses can be systematically increased by
$\sim$0.1 dex or decreased by $\sim$0.45 dex, thus resulting in a overall
rescaling of the stellar mass. A zero-point shift to the stellar mass scale
would not affect any of our environment analyses, which are
explicitly differential. 

The average error on the mass evaluation is $\sim$20\% and turns out
to be 10-15\% and 30-35\% for galaxies with probabilities greater and
less than 0.5 of being supercluster members, respectively.

\subsection{Galaxy Stellar Mass Function}
\label{sec:52} 

Figure~\ref{GSMF} shows the stellar mass function of galaxies with
$10^{8.75}$\,M$_\odot$ $< \mathcal{M} <$
10$^{12}$\,M$_\odot$. Based on our $K$-band completeness
limit of $K=18$, we conservatively derive the SMF down to
$\mathcal{M}$=10$^{8.75}$M$_\odot$, which corresponds to the estimated
stellar mass of the quiescent galaxy population at our $K$-band limit.
Fitting a single Schechter function, the recovered parameters are
log$_{10}$($\mathcal{M}^\star$)=11.16 and $\alpha$=--1.20
whose uncertainties are given by the confidence contours in Fig.~\ref{GSMF}.

We check the robustness of our result against possible completeness
issues of the galaxy sample used in deriving the SMF. According to
their location in the $R-K$ vs. $K$ colour-magnitude diagram, only 5\%
of the galaxies in the $K$ band catalogue which are possible
supercluster members (i.e. which lie along or below the red sequence)
do not have available estimates of probability. These galaxies are
uniformly distributed in the magnitude bins, and mostly lack optical
data (and hence probabilities) due to their location near bright
stars, which produce large ghosts in the optical data. We note also
that the low-mass galaxies we may have lost would contribute to
increase the low-mass-end slope of the SMF.

For what concerns the observed trend of the galaxy SMF, the slope is
in agreement with previous works concerning the SMF of field galaxies
(Cole et al. \citeyear{CNB01}, in the mass range
$10^{9}$--$10^{11.5}$\,M$_\odot$; Panter et al. \citeyear{pan04}, in
the mass range $10^{7.5}$--$10^{12}$\,M$_\odot$). On the other hand
the faint-end slope of the SSC turns out to be shallower than the
value given by Balogh et al. (\citeyear{BCZ01}) for the local mass
function of cluster galaxies [2MASS/ Las Campanas Redshift Survey
(LCRS)] although the latter is affected by a large uncertainty
($\Delta\alpha \sim$0.1 at 1$\sigma$). The value of
$\mathcal{M}^\star$ is higher than that observed in the field
(e.g. Bell et al. \citeyear{BMK03}), as expected for cluster
environment. The SMF of supercluster galaxies is characterised by an
excess of massive galaxies that is associated to the cluster BCGs.

In Fig.~\ref{MFenv} we show the SMF for the different supercluster
environments. Unlikely in the case of the LF no environmental trend is
seen in the slope of the SMFs (see Fig.~\ref{MFcont}). On the other
hand, the $\mathcal{M}^\star$ increase from low- to high-density
regions and the excess of high-mass galaxies remains dependent on the
environment. We remark that the uncertainty due to the adopted IMF
(see Sect.~\ref{sec:51}) does not affect our analysis since the
stellar masses at a fixed epoch will be all changed by the same amount.

\section{Discussion}
\label{sec:6}

\subsection{The faint-end upturn in the $K$-band LFs}
\label{sec:61}

We observed a steep faint-end slope for the Shapley supercluster
$K$-band LF (Fig.~\ref{totLF}) obtaining $\alpha$=--1.42, similarly to
values observed for the cluster LFs of Mobasher \& Trentham
(\citeyear{MT98}) and Andreon \& Pell\'o (\citeyear{AP00}).  De
Propris et al. (\citeyear{DES98}) observed a steeper slope of
$\alpha$=--1.7 deriving the NIR LF for a wider area
(29$^\prime$.2$\times$22$^\prime$.5) of the Coma cluster
(corresponding to $\sim$15$^\prime$$\times$11$^\prime$ at z$\sim$0.05)
down to $H$=16\,mag ($K\sim$15.7), but the authors conservatively
consider this result as an indication of a steep LF for dwarf galaxies
rather than a precise estimate of the slope.  On the other hand, other
studies found a less steep or flat faint-end slope in the NIR LF for
cluster galaxies (e.g. Lin et al. \citeyear{LMS04}).

The steep slope of the cluster LF has been discussed considering the
otherwise flat LF of field galaxies (e.g. Cole et
al. \citeyear{CNB01}) and how the environmental effects could shape the
galaxy LF. Balogh et al. (\citeyear{BCZ01}; see also Croton et
al. \citeyear{CFN05}) demonstrated that there is a statistical
difference between the cluster and field galaxy NIR LFs, with a
brighter M$^\star$ and a steeper faint-end slope for clusters with
respect to the field. An interpretation can be inferred from the
trends observed for the LFs of early- and late-type field
galaxies. Huang et al. (\citeyear{HGC03}) found that the later type
galaxies have a fainter M$^\star$ and a steeper slope with respect to
the early-type galaxies in the field. Their faint-end slope for
late-type galaxies is equal to the value we obtained for the SSC
($\alpha=$--1.42), although for a brighter sample. This suggests a
scenario whereby the faint-end upturn in the cluster LF is due to
late-type objects accreted from the field. This scenario is further
supported by the result of Harsono \& De Propris (\citeyear{har07})
who did not detect the up-turn in two intermediate redshift clusters
indicating that this feature has recent origins.

By considering the LFs of the Shapley supercluster in different
density environments, we found a steepening of the faint-end slope
which changes from $\alpha$=--1.33 to $\alpha$=--1.49, being
inconsistent at 2$\sigma$ c.l. between high- and low-density
regions. Moreover, the general shape of the LFs in the low-density
region turns out to be better reproduced using the combination of G+S
functions.  This observed bimodality in the LF and its variation with
environment suggest a scenario where bright and faint galaxy
populations have followed different evolution histories and indicate
that an environmental effect such as galaxy harassment and/or ram
pressure stripping could be responsible for shaping the LF. This
entails that the environment is responsible of the final mixture of the
galaxy types, in particular for the faint/low-mass galaxies.

The stellar and/or field contamination can artificially produce the
faint-end upturn. We checked carefully the issues of the stellar
contamination (see Sec.\,\ref{sec:32}) and we chose a control field
characterized by a deeper limiting magnitude and a larger area respect
to the Shapley $K$-band survey in order to rely on the star/galaxy
classification and to account for the field-to-field variance.  Assuming the
trend we observed to be real, we make a simple exercise to understand
how the survey depth can affect the measured shape of the LF. In
Fig.~\ref{LFcuts} we show that the slope of the faint-end becomes
clearly steeper as the depth of the sample is increased in all the
supercluster environments, demonstrating the need for such deep
data-sets to understand the role of environment on galaxy evolution.

\begin{figure*}
\centerline{{\resizebox{8cm}{!}{\includegraphics[angle=0]{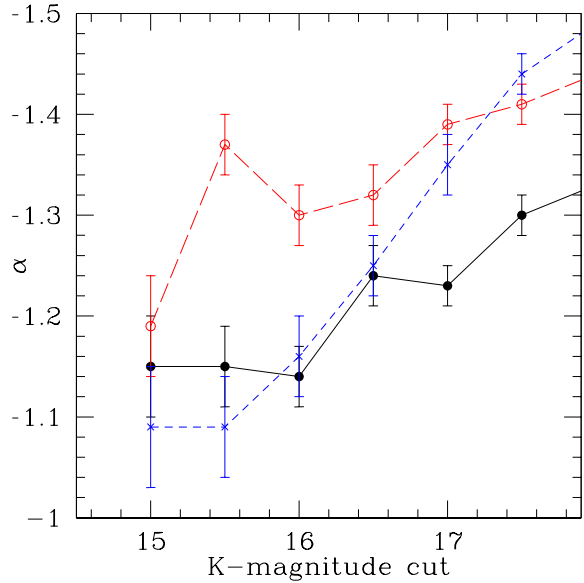}}}}     
\caption[]{Values of the LF slope $\alpha$ as function of the
different magnitude cuts in high- (continuous line and filled
circles), intermediate-(long dashed line and open circles) and
low-density (short dashed line and open circles) environments.}

      \label{LFcuts}

   \end{figure*}

The observed trend with environment is more dramatic for the Shapley
optical LFs obtained by MMH06. In particular, the faint-end slope
becomes steeper at $>3\sigma$ significance level from high-
($\alpha_{\mathrm{B}}$ =--1.46, $\alpha_{\mathrm{R}}$ =--1.30) to
low-density environments ($\alpha_{\mathrm{B}}$ =--1.66,
$\alpha_{\mathrm{R}}$ =--1.80) in $B$ and $R$ bands. Also the
bimodality of the galaxy LF, commonly observed for rich clusters
(e.g., Yagi et al. \citeyear{yag02}; Mercurio et
al. \citeyear{mer03}), turns out to be more evident in the optical
bands. We note that the SOS is about 1.5 magnitude deeper in $R$
band respect to the $K$ band considering $R-K$=2.6.

\subsection{The galaxy stellar mass function}
\label{sec:62}

The stellar mass functions derived for the supercluster galaxies are
in general agreement with those obtained by Baldry et
al. (\citeyear{BBB06}) who using SDSS data did not find changes in the
slope of the SMF with environment, except for changes in the
characteristic mass which increases with the local density. We do find
a steeper slope: $\alpha$=--1.20$\pm$0.02, instead of
$\alpha\simeq$--1 of Baldry et al. (\citeyear{BBB06}) who did not
quote error estimates. This discrepancy might be due to systematic
uncertainties in the mass derivation: in fact, using the MOPED
algorithm, Panter et al. (\citeyear{pan04}) obtained for the SDSS data
$\alpha$=--1.159$\pm$0.008 which is consistent within the errors with
our findings.  Another source of discrepancy can be due to the use of
the $K$-band data to estimate the galaxy stellar masses, but the slope
obtained by Bell et al. (\citeyear{BMK03}) combining 2MASS and SDSS
data-sets is significantly lower ($\alpha$=--0.86) respect to those of
SSC in all environments. As shown by Bell et
al. (\citeyear{BMK03}) the difference here may be due to the fact that
2MASS misses low surface brightness galaxies. They estimated the
$K$-band flux and stellar masses from the optical photometry for the
galaxies not detected in $K$-band and produced a $g$-band
derived stellar mass function with $\alpha$=--1.10.

\citet{BGD08} found a strong low-mass upturn below
$\mathcal{M}=10^9$M$_\odot$ analysing the NYU-VAGC\footnote{New York
University Value-Added Galaxy Catalog (Blanton et
al. \citeyear{bla05a}, \citeyear{bla05b})}, their SMF being fitted by
a S+S function with, in the low-mass end, $\alpha$=--1.58$\pm$0.02.
An upturn in the SMF was also shown by Jenkins et
al. (\citeyear{JHM07}) for the Coma cluster by means of {\it Spitzer}
InfraRed Array Camera (IRAC) observations. They observed a steep
increase in two different Coma regions below
$\mathcal{M}=10^{8.2}$M$_\odot$ with $\alpha\sim$--2 (but they did not
quantify the trend with a fit of the data).

The fact that we do not observe any upturn in the SMF of the Shapley
supercluster can be related to the different mass range or to
environmental differences. Jenkins et al. (\citeyear{JHM07}) and
\citet{BGD08} extend their analysis to
$\mathcal{M}$=10$^{7.5}$M$_\odot$ and $\mathcal{M}$=10$^{8}$M$_\odot$,
respectively, with the faint-end upturn becoming evident at
$\mathcal{M}<$10$^{9}$M$_\odot$, very close to our mass limit is
$\mathcal{M}$=10$^{8.75}$M$_\odot$.  This suggest that the reason why
we see an upturn in the $K$-band LF, but not in the SMF, is that the
galaxies which cause the faint-end upturn are star-forming galaxies
with low stellar mass-to-light ratios and stellar masses in the range
10$^{8}<\mathcal{M}<$10$^{9}$M$_\odot$. This would be consistent with
Bolzonella et al. (\citeyear{bol09}) who found that the upturn was due
to late-type galaxies, but inconsistent with the finding of MMH06 in
which the upturn appeared due to red sequence galaxies.

Jenkins et al. (\citeyear{JHM07}) analysed two regions of the Coma
cluster: one in the cluster centre (0.733\,deg$^2$
corresponding to $\sim$2.1\,Mpc$^2$) and one off-centre region
(0.555\,deg$^2$ corresponding to $\sim$1.6\,Mpc$^2$)
located $\sim$1.7 Mpc southwest. The area of the SSC covered both by
the SOS and the $K$-band survey is $\sim$30\,Mpc$^2$. Therefore, we
are studying a much larger area that comprises infall regions as well as cluster cores. Dynamical analysis indicates that at least a region
of radius 11\,Mpc centred on the central cluster
A\,3558, and possibly the entire supercluster, is past turnaround and
is collapsing (Reisenegger et al. \citeyear{rei00}), while the core
complex itself is in the final stages of collapse, with infall
velocities reaching $\sim$2000\,km\,s$^{-1}$. This differences in the
environments may be responsible of differences in the stellar mass
function for the two cosmic structures. Another, possible issue can be
related to the background subtraction (see Rines \& Geller \citeyear{rin08}).

Investigating the environmental effect on the SMF in the SSC, we find
that both $\mathcal{M}^\star$ and the excess of galaxies at the
bright-end increase as foreseen by the hierarchical models where the
most massive galaxies formed in the density peaks. On the other hand
the faint-end slope does not change in the different supercluster
environments suggesting that the mechanism that is acting in shaping
the LF (see Sect.~\ref{sec:43}) does not significantly affect the galaxy
stellar masses.

\subsection{Comparison with the simulations}
\label{sec:63}

\begin{figure}
\centerline{{\resizebox{10cm}{!}{\includegraphics[angle=0]{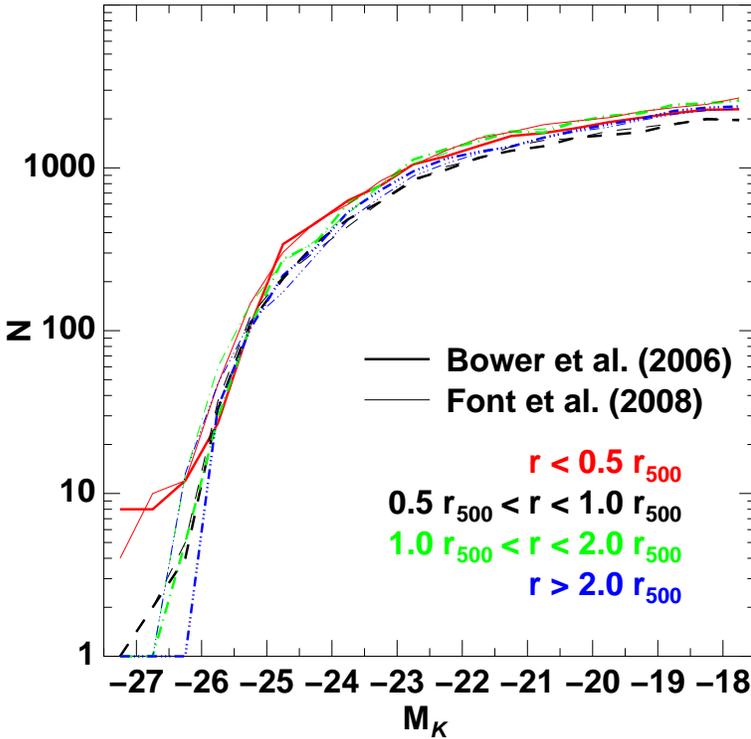}}}}
\caption{Composite $K$-band LFs for the 20 most massive clusters in the
Millennium simulation, based on the semi-analytic models of Bower et
al. \citeyear{BBM06} (thick lines) and Font et al. \citeyear{Font08}
(thin lines), in bins of projected cluster-centric radius. The four
different curves correspond to $r<0.5\,r_{500}$ (solid lines),
$0.5<r<1.0\,r_{500}$ (dashed lines), $1.0<r<2.0\,r_{500}$ (dot-dashed
lines) and $r>2.0\,r_{500}$ (dot-dot-dashed lines).}
\label{bower}
\end{figure}

We should expect changes in the $K$-band luminosity function with
environment to reflect two processes: (i) the effect of the diverse
merging histories of galaxies in different environments within the
context of the hierarchical merging scenario, in which cluster galaxies
are likely to have formed earlier and had a more active merger history
than field galaxies that form in the smoother low-density regions; and
(ii) the later impact of environmental processes such as tidal
stripping, which may drastically reduce the stellar mass of galaxies
in high-density regions. While it is very difficult to quantify the
latter's contribution to the galaxy luminosity/mass function, we can
attempt to measure the former contribution by comparison to the
cosmological numerical simulations.

To this aim, we extracted galaxy catalogues in the vicinity of the 20
most massive dark matter halos from the Millennium simulation
(Springel et al. \citeyear{Spr05}), corresponding to galaxy clusters
with masses 7--2$3\times10^{14}$M$_{\odot}$ and velocity dispersions
($800<\sigma<1400$\,km\,s$^{-1}$). These simulations cover a
(500$h^{-1}$Mpc)$^{3}$ volume, producing DM halo and galaxy catalogues
based on the semi-analytic models (SAMs) of Bower et
al. (\citeyear{BBM06}) and Font et al. (\citeyear{Font08}) for which
positions, peculiar velocities, absolute magnitudes and halo masses
are all provided at 63 snapshots to $z=0$, allowing the orbit of each
galaxy with respect to the cluster centre to be followed. We select
member galaxies from these twenty clusters that have $M_{K}<-17.5$,
and lie within 5\,Mpc of the cluster centre, and show in
Fig.~\ref{bower} the stacked $K$-band cluster luminosity functions of
galaxies as a function of projected cluster-centric distance, scaled
by the $r_{500}$ value of each cluster's DM halo. We show the LFs
obtained from both the Bower et al. \citeyear{BBM06} (thick lines) and
Font et al. \citeyear{Font08} (thin lines) SAM galaxy catalogues, and
the four different curves correspond to $r<0.5\,r_{500}$,
$0.5<r<1.0\,r_{500}$, $1.0<r<2.0\,r_{500}$ and $r>2.0\,r_{500}$.  For
comparison, the $r_{500}$ values for the clusters in the SSC derived
from Chandra X-ray observations (Sanderson et al. \citeyear{SPO06};
Haines et al.  \citeyear{HSE09}) are 1.29\,Mpc (22.9\,arcmin at
$z=0.048$) for A\,3558, 0.91\,Mpc (16.1\,arcmin) for A\,3562 and
SC\,1327-312 and 0.76\,Mpc (13.5\,arcmin) for SC\,1329-317. By
inspection of Fig.~1, our high-, intermediate- and low-density regions
correspond approximately to the $r<0.5\,r_{500}$, $0.5<r<1.0\,r_{500}$
and $1.0<r<2.0\,r_{500}$ radial bins respectively.

It is immediately apparent from Fig.~\ref{bower} that there is little
if any environmental dependence of the $K$-band luminosity function,
except at the very bright end ($M_K<-26$), while equally there are no
significant differences between the two semi-analytic models. This is
confirmed if we compare the derived best-fitting single Schechter functions to
the SAM LFs, with $M^\star_{K}=-24.42\pm0.04$, $\alpha=-1.134\pm0.007$ for
$r<0.5\,r_{500}$ and $M^\star_{K}=-24.37\pm0.03$, $\alpha=-1.167\pm0.003$
for $r>2.0\,r_{500}$. Similarly, if we select just those galaxies which
have already passed through the cluster core (based on following their
orbits), and hence are most likely to correspond to the red sequence
population, we find no significant variation in the LFs with
cluster-centric radius. There is also no sign of an upturn at faint
magnitudes, although we note the SAM galaxy catalogues start to become
incomplete at $M_{K}{\gtrsim}-20$ due to the limited mass resolution
in the Millennium simulation (Bower et al. \citeyear{BBM06}). The only
clear difference among the LFs is the excess at the extreme bright end
($M_{K}<K^{*}-2$) in the inner cluster region, due to the BCG
population. Moreover, the $\chi^2$ test gives a probability of
0\% that the observed and SAM LFs are drawn by the same parent
distribution.

Although the merging history of galaxies at the faint-end
($M_{K}>-20$) may not be fully resolved by the semi-analytic models,
we may expect at least their stellar masses (and hence $K$-band
luminosities) to be reasonably robust (Bower et
al. \citeyear{BBM06}). In previous works discussing the presence of a
dip in the galaxy luminosity function, it has been suggested the
merging of $~L^{*}$ galaxies as a
possible pathway to form the dip. However the consistency of the
luminosity functions with cluster-centric radius, suggests that the
variations observed in the Shapley supercluster (Fig.~6) are not
primarily due to variations in the merging histories of galaxies, this
being a process that should be well described by the semi-analytic
models. Instead processes such as tidal stripping or harassment of
infalling spiral galaxies, which are not yet included in the
semi-analytic models seem more plausible pathways to produce the
observed variations in the galaxy LFs.

\section{Summary and conclusions}
\label{sec:7}

It is well known that the NIR luminosities provide a more reliable
estimate of the stellar masses compared to the optical ones due to the
fact that the mass-to-light ratios in the NIR vary by at most a factor
2 across a wide range of star formation histories (Bell \& de Jong
\citeyear{BdJ01}) in comparison to the much larger variations of the
M/L ratios (up to a factor 10) observed at optical wavelengths. In
addition the effects of the extinction are much weaker at infrared
wavelengths that in the optical ones, and k-corrections for infrared
colours are only weakly dependent on the Hubble type and vary slowly
with redshift (e.g. Poggianti \citeyear{pog97}).  In our study we 
exploit new deep ($K$=18\,mag) $K$-band imaging of the SSC
complemented by the deep optical imaging down to $B$=22.5\,mag and
$R$=22\,mag from the SOS, and a spectroscopically confirmed
supercluster sample of $\sim$650 galaxies across the same field which
is $\sim$90\% complete at $R<$16\,mag. We present an analysis of the
$K$-band LF of galaxies as a function environment and we derive the
galaxy SMF in order to constrain the physical mechanisms that
transform galaxies in different environments as function of galaxy
mass.

Our results are summarized as follows.

\begin{description}

\item[-] The $K$-band LF can be fitted by a single Schechter function,
with $M_{K}^{*}=-24.96\pm0.10$ and $\alpha=-1.42\pm0.03$ in agreement
with previous works of comparable depth.

\item[-] The $K$-band LF faint-end slope becomes steeper from high- to
low-density environments varying from --1.33 to --1.49, being
inconsistent at the 2$\sigma$ c.l. (see Tab~\ref{fitsLF}),
indicating that the faint galaxy population increases in low-density
environments. Such an environmental dependence confirms our finding for
the optical LFs derived in the same supercluster regions although the
changes in slope are less dramatic at NIR wavebands.

\item[-] The observed trend of the galaxy SMF presents a slope in
agreement with previous works concerning the SMF of field galaxies
(Cole et al. \citeyear{CNB01}; Panter et al. \citeyear{pan04}).  The
value of $\mathcal{M}^\star$ is higher with respect to that observed
in the field (e.g. Bell et al. \citeyear{BMK03}), as expected for
cluster environment. The SMF of supercluster galaxies is characterised
by an excess of massive galaxies that is associated to the cluster
BCGs.  Discrepancies with previous work that observed a strong
faint-end upturn (Baldry et al. \citeyear{BGD08} and Jenkins et
al. \citeyear{JHM07}) can be related to the different mass ranges
investigated and/or environmental differences in the analysed
structures.

\item[-] Differently from the LF no environment effect is found in the
slope of the SMFs. On the other hand, the $\mathcal{M}^\star$ increase
from low- to high-density regions and the excess of galaxies at the
bright-end is also dependent on the environment. This trend is in
general agreement with the results of Baldry et al. (\citeyear{BBB06}).

\end{description}

In order to interpret our findings, we use the Millennium
simulation which produce DM halo and galaxy catalogues based on
SAMs. The cluster NIR LFs obtained using the simulated catalogues do
not show any significant variation with cluster-centric radius, thus
suggesting that the variations observed in the LFs of the Shapley
supercluster are not driven by variations in the merging histories of
the galaxies, but are likely related to processes such as tidal
stripping or harassment of infalling galaxies.

By comparing the effect of environment at optical and NIR wavebands in
shaping the LFs and taking into account that the slope of the galaxy
SMF is invariant with respect to the environment, it seems that the
physical mechanism responsible for the transformation of galaxies
properties in different environment are related to the quenching of the
star formation rather than mass-loss. 

This suggests that the mechanism responsible for shaping the LF and
SMF is partially related to the mass loss due to tidal stripping or
galaxy harassment, but gas stripping by the ICM can also affect the
galaxy population removing the cold gas supply and thus rapidly
terminating ongoing star-formation. These mechanisms all require a
dense ICM and so their evolutionary effects on massive galaxies are
limited to the cores of clusters, but can extend to poorer
environments for dwarf galaxies which are easier to strip. The
infalling galaxies are probably late-type (see MMH06) that are
affected by gas stripping entering in the supercluster.  On the other
hand, the depth of the NIR survey could affect the present results by
not allowing us to detect an upturn of the SMF which can be detected
at a mass range lower than reached here (see Jenkins et
al. \citeyear{JHM07}). In order to identify the mechanisms which
drive galaxy evolution in the supercluster environment we are
undertaking a survey with the integral field spectrograph WiFeS which
will provide a unique data set to investigate in details the stellar
populations and kinematics for a subsample of the Shapley galaxies. 

\section*{Acknowledgements}

This work was carried out in the framework of the collaboration of the
FP7-PEOPLE-IRSES-2008 project ACCESS. AM acknowledges financial
support from INAF - Osservatorio di Capodimonte. CPH acknowledges
financial support from STFC. RJS was supported under STFC rolling
grant PP/C501568/1 ``Extragalactic Astronomy and Cosmology at Durham
2005-2010''. The authors wish to thank the staff at UKIRT, CASU and
WFAU for making the observations and processing the data, and the
UKIDSS project and archive which has been used for computing the field
galaxy counts (3rd data release). The authors thank the anonymous
referee for helping to improve this work.

\label{lastpage}

\end{document}